\documentclass[12pt, twoside]{article}

\usepackage{a4wide, amsmath, latexsym, epsfig,
  psfrag,graphicx, rotating, fancyhdr, tabularx, afterpage}

\usepackage{feynarts}
\usepackage{axodraw}

\newcommand{\id}{{\rm 1\kern-.12em
\rule{0.3pt}{1.5ex}\raisebox{0.0ex}{\rule{0.1em}{0.3pt}}}}

\newcommand{\seff}{\sin^2\!\theta_{\rm eff}}
\newcommand{\Li}{\mathrm{Li}_2}

\begin{document}

\thispagestyle{empty}
\setcounter{page}{0}
\def\thefootnote{\fnsymbol{footnote}}

{\textwidth 15cm

\begin{flushright}
MPP-2006-134\\
hep-ph/0610312 \\
\end{flushright}

\vspace{2cm}

\begin{center}

{\Large\sc {\bf The effective electroweak mixing angle 
                \boldmath{$\sin^2\theta_{\rm eff}$} \\[0.3cm]
               with two-loop bosonic contributions}}

\vspace{2cm}

\sc{W. Hollik$^{1}$, U. Meier$^{1}$}\rm and \sc{S. Uccirati$^{2}$}\rm
\vspace{1cm}

     $^{1}$ Max-Planck-Institut f\"ur Physik \\
     (Werner-Heisenberg-Institut)\\
     D-80805 M\"unchen, Germany

\vspace*{0.4cm}

     $^{2}$ Dip.\ di Fisica Teorica, Universit\`a di Torino, \\
       INFN Sezione di Torino\\
       10125 Torino, Italy

\end{center}

\vspace*{2cm}

\begin{abstract}
\noindent
We present the results for the 
full electroweak two-loop bosonic contributions
to the effective leptonic mixing angle of the $Z$ boson,
$\sin^2\theta_{\rm eff}$, in the Standard Model.
A method applied 
to extract collinear divergences from two-loop vertex-functions is 
described.
Comparisons of our results 
with those from a recent previous calculation 
show complete agreement. 
\end{abstract}

}
\def\thefootnote{\arabic{footnote}}
\setcounter{footnote}{0}

\newpage

\section{Introduction}

Measurements of the various asymmetries of the $Z$ resonance have determined 
the effective leptonic mixing angle $\seff$ with high accuracy. 
The current experimental value is 
$0.23153 \pm 0.00016$~\cite{unknown:2004qh}; a linear electron-positron
collider with GigaZ capabilities could even reach an accuracy of $1.3 \times
10^{-5}$~\cite{Aguilar-Saavedra:2001rg, Baur:2001yp}. Comparison with the
theoretical prediction of the Standard Model 
yields stringent bounds on the Higgs-boson mass $M_H$.
The importance and precision of $\seff$ requires sufficient control
on the theoretical accuracy of the prediction
by providing adequate higher-order calculations.

$\seff$ is determined from the ratio of the dressed vector and 
axial vector couplings 
$g_{V,A}$ of the $Z$ boson to leptons~\cite{Bardin:1997xq},
\begin{eqnarray}
\seff &=& \frac{1}{4}  \left( 1-\mathrm{Re}\, \frac{g_V}{g_A} \right) . \label{s2w}
\end{eqnarray}
In the on-shell renormalization scheme, 
it is related to the vector-boson mass ratio or,
equivalently, to the on-shell quantity $s_W^2=1-M_W^2/M_Z^2$
via
\begin{eqnarray}
\seff = \kappa\,s_W^2 \, , \qquad\quad   \kappa = 1+\Delta\kappa \, , 
\label{kappa}
\end{eqnarray}
involving the $\kappa$ factor, 
which is unity at the tree level and accommodates the 
higher-order contributions in $\Delta\kappa$. 
$M_W$ can be related to the precisely known Fermi constant $G_{\mu}$ 
via the relation
\begin{eqnarray} \label{interdependence}
M_W^2 \left( 1-\frac{M_W^2}{M_Z^2} \right) &=& 
       \frac{\pi\alpha}{\sqrt{2} G_F} (1+\Delta r) \, , 
\end{eqnarray} 
where $\Delta r$ summarizes the higher-order contributions. 
Beyond the one-loop order,
the universal QCD corrections~\cite{Djouadi:1987gn,Chetyrkin:1995js}, 
the complete electroweak fermionic~\cite{Freitas:2002ja} 
and bosonic~\cite{Awramik:2002wn} two-loop corrections, 
as well as leading higher-order 
contributions via the $\rho$-parameter~\cite{Faisst:2003px,Boughezal:2004} 
and the $S$-parameter~\cite{Boughezal:2005eb} are known for the Standard Model.

The universal higher-order terms to the self-energies from QCD and 
through the $\rho$- and $S$-parameter enter also 
the quantity $\Delta\kappa$ in (\ref{kappa}).
The fermionic electroweak two-loop corrections 
(involving at least one closed fermion loop) 
have been established~\cite{Awramik:2004ge,Hollik:2005va}; 
the $M_H$-dependence of the bosonic corrections
to $\Delta\kappa$
was given in~\cite{Hollik:2005ns}, 
and only recently a complete calculation of the
bosonic contributions was reported~\cite{Awramik:2006ar}.

In this paper we complete
our result for the bosonic two-loop contributions
by  accomplishing the $M_H$-independent part of $\Delta\kappa$, which was
missing in~\cite{Hollik:2005ns}.  
Together with our result from~\cite{Hollik:2005ns}, this 
yields an independent result for the complete 
bosonic two-loop corrections for $\seff$, based on different methods.

\section{Calculational strategy}
\label{2loop}
Expanding the dressed couplings in (\ref{s2w}) according to
$g_{V,A} = g_{V,A}^{(0)} \left(1+g_{V,A}^{(1)}+g_{V,A}^{(2)}+\cdots\right)$ 
in powers of $\alpha$, the $O\left(\alpha^2\right)$ 
contribution to $\seff$ in the loop expansion 
\begin{eqnarray}
\seff &=&
\seff^{(0)}+\seff^{(1)}+
\seff^{(2)}+\mathcal{O}\left(\alpha^3\right)
\end{eqnarray}
is obtained,
\begin{eqnarray}
\seff^{(2)} &=& -\frac{g_V^{(0)}}{4 g_A^{(0)}} \;
\mathrm{Re}\left(g_V^{(2)}-g_A^{(2)}+g_A^{(1)} 
\left( g_A^{(1)}-g_V^{(1)} \right)\right).\label{expand}
\end{eqnarray}
Hence, besides two-loop diagrams, also
products of one-loop contributions have to be taken into account. They play an
important role in the cancellation of IR-divergences (see
section~\ref{IR}). But the major task consists in the calculation of
irreducible two-loop $Z\ell\ell$-vertex diagrams
($\ell = e$, to be precise). 
The $Z$-boson couplings in (\ref{s2w}) appear in the renormalized 
$Z\ell\ell$ vertex for on-shell $Z$ bosons; 
for the two-loop contributions entering the
expression~(\ref{expand}) we need the renormalized two-loop vertex 
\begin{eqnarray}
\hat{\Gamma}_\mu^{Z\ell\ell\,(2)}(M^2_Z) &=& \gamma_\mu\,
   \left( g_V^{(2)} - g_A^{(2)} \gamma_5 \right) \, . 
\end{eqnarray} 
As done in~\cite{Hollik:2005va,Hollik:2005ns} we split the renormalized vertex into two $UV$-finite pieces,
\begin{eqnarray}
\hat{\Gamma}_\mu^{Z\ell\ell\,(2)}(M^2_Z) &=& 
\Gamma_\mu^{Z\ell\ell\,(2)} (M^2_Z) + \Gamma_\mu^{CT}
\nonumber \\[0.2cm]
 &=&  
\left[ \Gamma_\mu^{Z\ell\ell\, (2)}(0)+ \Gamma_\mu^{CT}\right]
+\left[ \Gamma_\mu^{Z\ell\ell\,(2)}(M^2_Z) - 
\Gamma_\mu^{Z\ell\ell\,(2)}(0) \right] .\label{split}
\end{eqnarray}
$\Gamma_\mu^{Z\ell\ell\,(2)}\left(M_Z^2\right)$ 
denotes the corresponding unrenormalized  $Z\ell\ell$ vertex
for on-shell leptons and momentum transfer $P^2=M_Z^2$,
and $\Gamma_\mu^{CT}$ is the two-loop counter term.
Details on the renormalization are given in~\cite{Hollik:2005va}, where the 
expressions are general and comprise also the bosonic two-loop contributions.

The first term in (\ref{split}) can be computed as in the fermionic case, 
which means generating
Feynman diagrams with the help of
\it{FeynArts}\rm~\cite{Hahn:2000jm} and applying 
\it{TwoCalc}\rm~\cite{Weiglein:1993hd} to reduce the amplitudes to standard
integrals. The resulting vacuum integrals are calculated using analytic
results \cite{'tHooft:1978xw,Davydychev:1992mt}, whereas the two-loop
self-energies with non-vanishing external momenta,
which appear in the counter terms, 
are calculated with the help of one-dimensional integral representations~\cite{Bauberger:1994nk} or the methods described in~\cite{Ferroglia:2003yj}.

The $IR$-finite contributions to the second term in (\ref{split}) can be
calculated in analogy to the $M_H$-dependent part of the bosonic
corrections~\cite{Hollik:2005ns}. This means that vertex corrections are
computed applying the methods from~\cite{Ferroglia:2003yj} together with further
improvements in order to increase the numerical stability. Non-planar diagrams
are calculated using the method described in~\cite{Hollik:2005ns}, and
diagrams with self-energy insertions are computed using dispersion relations as described in~\cite{Hollik:2005va}.

\section{Treatment of IR- and collinear divergences}\label{IR}

\begin{figure}[!htb]
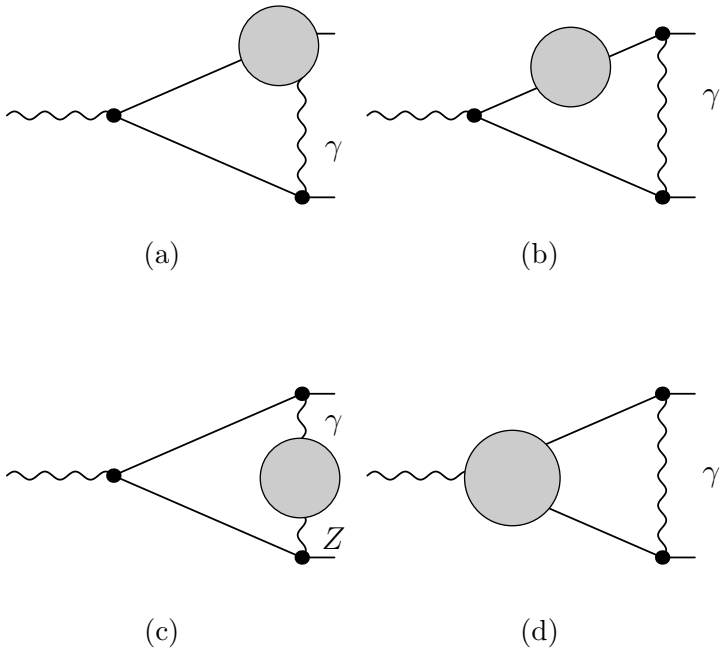

\vspace*{-2 cm}
\begin{feynartspicture}(450,300)(2,2.2)

\FADiagram{(a)}
\FAProp(0.,10.)(6.5,10.)(0.,){/Sine}{0}
\FAProp(20.,15.)(18.,15.)(0.,){/Straight}{0}
\FAProp(20.,5.)(18.,5.)(0.,){/Straight}{0}
\FAProp(6.5,10.)(18.,15.)(0.,){/Straight}{0}
\FAProp(6.5,10.)(18.,5.)(0.,){/Straight}{0}
\FAProp(18.,15.)(18.,5.)(0.,){/Sine}{0}
\FAVert(6.5,10.){0}
\FAVert(18.,15.){0}
\FAVert(18.,5.){0}
\FALabel(20., 7.)[t]{$\gamma$}

\FADiagram{(b)}

\FAProp(0.,10.)(6.5,10.)(0.,){/Sine}{0}
\FAProp(20.,15.)(18.,15.)(0.,){/Straight}{0}
\FAProp(20.,5.)(18.,5.)(0.,){/Straight}{0}
\FAProp(6.5,10.)(18.,15.)(0.,){/Straight}{0}
\FAProp(6.5,10.)(18.,5.)(0.,){/Straight}{0}
\FAProp(18.,15.)(18.,5.)(0.,){/Sine}{0}
\FAVert(6.5,10.){0}
\FAVert(18.,15.){0}
\FAVert(18.,5.){0}
\FALabel(20., 10.)[t]{$\gamma$}

\FADiagram{(c)}

\FAProp(0.,10.)(6.5,10.)(0.,){/Sine}{0}
\FAProp(20.,15.)(18.,15.)(0.,){/Straight}{0}
\FAProp(20.,5.)(18.,5.)(0.,){/Straight}{0}
\FAProp(6.5,10.)(18.,15.)(0.,){/Straight}{0}
\FAProp(6.5,10.)(18.,5.)(0.,){/Straight}{0}
\FAProp(18.,15.)(18.,5.)(0.,){/Sine}{0}
\FAVert(6.5,10.){0}
\FAVert(18.,15.){0}
\FAVert(18.,5.){0}
\FALabel(20., 13.)[t]{$\gamma$}
\FALabel(20., 6.5)[t]{$Z$}

\FADiagram{(d)}

\FAProp(0.,10.)(6.5,10.)(0.,){/Sine}{0}
\FAProp(20.,15.)(18.,15.)(0.,){/Straight}{0}
\FAProp(20.,5.)(18.,5.)(0.,){/Straight}{0}
\FAProp(6.5,10.)(18.,15.)(0.,){/Straight}{0}
\FAProp(6.5,10.)(18.,5.)(0.,){/Straight}{0}
\FAProp(18.,15.)(18.,5.)(0.,){/Sine}{0}
\FAVert(6.5,10.){0}
\FAVert(18.,15.){0}
\FAVert(18.,5.){0}
\FALabel(20., 10.)[t]{$\gamma$}

\GCirc(197.,230.){15}{0.8}
\GCirc(307.,222.){15}{0.8}
\GCirc(205.,67.){15}{0.8}
\GCirc(285.,67.){18}{0.8}

\end{feynartspicture}
\caption{Diagrams with IR-divergences}
\label{diagbossoft}
\end{figure}

As a new feature compared to the $M_H$-dependent subset of the bosonic
corrections, diagrams with internal photons appear that can be IR-divergent. 
These divergences, however,
cancel in the complete result rendering $\seff$ as a IR-finite quantity.

\begin{figure}[!htb]
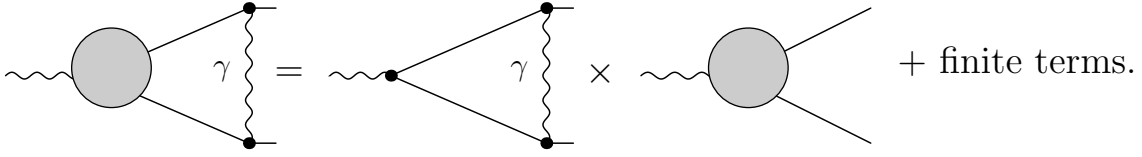

\vspace*{-4 cm}
\begin{feynartspicture}(450,300)(4,1.4)

\FADiagram{}

\FAProp(0.,10.)(6.5,10.)(0.,){/Sine}{0}
\FAProp(20.,15.)(18.,15.)(0.,){/Straight}{0}
\FAProp(20.,5.)(18.,5.)(0.,){/Straight}{0}
\FAProp(6.5,10.)(18.,15.)(0.,){/Straight}{0}
\FAProp(6.5,10.)(18.,5.)(0.,){/Straight}{0}
\FAProp(18.,15.)(18.,5.)(0.,){/Sine}{0}
\FAVert(6.5,10.){0}
\FAVert(18.,15.){0}
\FAVert(18.,5.){0}
\FALabel(16., 10.)[t]{$\gamma$}
\FALabel(21., 9.5)[t]{\large{$=$}}

\FADiagram{}
\FAProp(2.,10.)(6.5,10.)(0.,){/Sine}{0}
\FAProp(20.,15.)(18.,15.)(0.,){/Straight}{0}
\FAProp(20.,5.)(18.,5.)(0.,){/Straight}{0}
\FAProp(6.5,10.)(18.,15.)(0.,){/Straight}{0}
\FAProp(6.5,10.)(18.,5.)(0.,){/Straight}{0}
\FAProp(18.,15.)(18.,5.)(0.,){/Sine}{0}
\FAVert(6.5,10.){0}
\FAVert(18.,15.){0}
\FAVert(18.,5.){0}
\FALabel(16., 10.)[t]{$\gamma$}
\FALabel(22., 10.)[t]{\large{$\times$}}

\FADiagram{}

\FAProp(3.,10.)(10.,10.)(0.,){/Sine}{0}
\FAProp(10.,10.)(20.,15.)(0.,){/Straight}{0}
\FAProp(10.,10.)(20.,5.)(0.,){/Straight}{0}
\FAVert(10.,10.){0}

\FADiagram{}
\FALabel(0., 10.)[l]{\large{$+$ finite terms.}}

\GCirc(285.,130.){15.}{0.8}
\GCirc(45.,130.){15.}{0.8}

\end{feynartspicture}
\vspace*{-2 cm}
\caption{Compensation of IR-divergences}
\label{diagcancelsoft}
\end{figure}

The IR-divergent diagrams are shown in Fig.~\ref{diagbossoft}. The grey circles represent one-loop insertions.
In order to verify the cancellation of these divergences, we have used the
methods described in~\cite{Passarino:2006gv}. These methods allow to extract
all appearing IR-divergences in terms of IR-divergent one-loop
integrals. After this extraction the cancellation of these integrals can be
checked analytically. The cancellation occurs between the various
two-loop diagrams within the set of Fig.~\ref{diagbossoft}(a)-(c),   
and  between the two-loop diagrams of Fig.~\ref{diagbossoft}(d) 
and the product of one-loop diagrams occuring in (\ref{expand}). 
This is shown schematically in Fig.~\ref{diagcancelsoft}.

Moreover, collinear divergences appear. In order to regularize them we have
kept the electron mass $m_e$ finite where necessary. 
In addition we have expanded the resulting expressions in $m_e$, so that the
divergent behavior shows up as terms proportional to $\ln^2(m^2_e)$ and
$\ln(m^2_e)$. Afterwards it was checked analytically that the terms proportional to $\ln^2(m^2_e)$ cancel, whereas the cancellation of the $\ln(m^2_e)$-terms was checked numerically. Our methods to extract these logarithms are explained in the following section \ref{collinear} by means of the two non-planar diagrams shown in Fig.~\ref{boscoll}.

\section{Extraction of collinear divergences}\label{collinear}

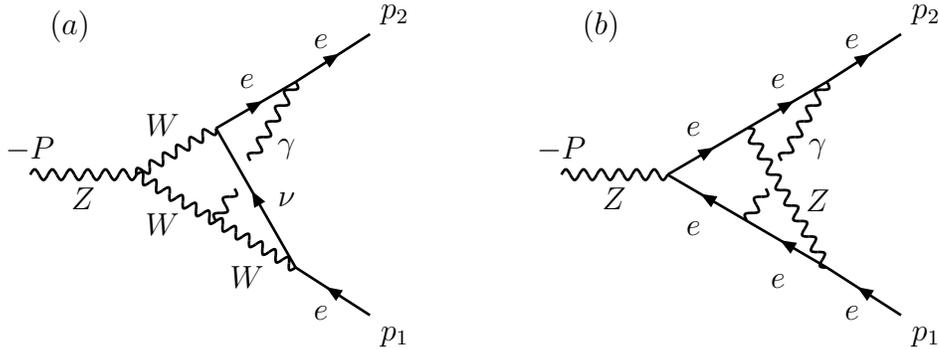
\begin{figure}[ht]
\begin{center}
\begin{picture}(150,75)(0,0)
 \SetWidth{1.}
 \Text(0,5)[cb]{$-P$}
 \Text(138,-65)[cb]{$p_1$}
 \Text(138,57)[cb]{$p_2$}            
 \Photon(0,0)(40,0){2}{6}            \Text(20,-13)[cb]{$Z$}
 \ArrowLine(128,-53)(100,-35)        \Text(110,-55)[cb]{$e$}
 \ArrowLine(100,35)(128,53)          \Text(110,48)[cb]{$e$}
 \Photon(70,17.5)(40,0){2}{6}        \Text(50,15)[cb]{$W$}
 \Photon(70,-17.5)(40,0){2}{6}       \Text(50,-23)[cb]{$W$}
 \ArrowLine(70,17.5)(100,35)         \Text(82,33)[cb]{$e$}
 \Photon(100,-35)(70,-17.5){2}{6}    \Text(82,-43)[cb]{$W$}
 \Photon(100,35)(83,5.25){2}{5}
 \Photon(77,-5.25)(70,-17.5){2}{2}   \Text(97,6)[cb]{$\gamma$}
 \ArrowLine(100,-35)(70,17.5)        \Text(97,-12)[cb]{$\nu$}
 \Text(15,50)[cb]{$(a)$}
\end{picture}
\qquad\qquad
\begin{picture}(150,75)(0,0)
 \SetWidth{1.}
 \Text(0,5)[cb]{$-P$}
 \Text(138,-65)[cb]{$p_1$}
 \Text(138,57)[cb]{$p_2$}            
 \Photon(0,0)(40,0){2}{6}            \Text(20,-13)[cb]{$Z$}
 \ArrowLine(128,-53)(100,-35)        \Text(110,-55)[cb]{$e$}
 \ArrowLine(100,35)(128,53)          \Text(110,48)[cb]{$e$}
 \ArrowLine(40,0)(70,17.5)           \Text(50,15)[cb]{$e$}
 \ArrowLine(70,-17.5)(40,0)          \Text(50,-23)[cb]{$e$}
 \ArrowLine(70,17.5)(100,35)         \Text(82,33)[cb]{$e$}
 \ArrowLine(100,-35)(70,-17.5)       \Text(82,-43)[cb]{$e$}
 \Photon(100,35)(83,5.25){2}{5}
 \Photon(77,-5.25)(70,-17.5){2}{2}   \Text(97,6)[cb]{$\gamma$}
 \Photon(100,-35)(70,17.5){2}{9}     \Text(97,-14)[cb]{$Z$}
 \Text(15,50)[cb]{$(b)$}
\end{picture}
\end{center}
\vspace{2cm}
\caption[]{Examples for collinearly divergent diagrams. 
All momenta are incoming, and $P= p_1+p_2$.}
\label{boscoll}
\end{figure}

Collinear divergences can show up in massive diagrams which do not contain any
soft divergences, or they can overlap with soft divergences. 
In both cases we have managed to extract explicitly the coefficients of the
logarithms in $m^2_e$ (which is the natural regulator for these divergences) and
the constant term. For overlapping divergences we have used the results of
\cite{Passarino:2006gv}, while for the other ones we have used a subtraction
method in parametric space. In order to explain it we consider the two
collinear configurations of the non-planar diagrams in Fig.~\ref{boscoll}, 
which arise in the computation.

\noindent
In momentum space the generic diagram contains the following scalar and tensor
integrals,
\begin{eqnarray}
V^{\left(1,\mu,\mu\nu,\mu\nu\rho\right)}_{222} &=& 
- \frac{1}{\pi^4} \int d^4 q_1 \int d^4 q_2 
  \frac{\left(1, q_1^\mu,q_1^\mu q_1^\nu,q_1^\mu q_1^\nu q_1^\rho\right)}
       {[1][2][3][4][5][6]},
\end {eqnarray}
\begin{eqnarray}
&&\nonumber 
[1]= q_1^2-m_1^2,\qquad[2]= (q_1-p_2)^2-m_2^2,\qquad[3]= (q_1-q_2+p_1)^2-m_3^2,\\
&&\nonumber 
[4]= (q_1-q_2-p_2)^2-m_4^2,\qquad[5]= q_2^2-m_5^2,\qquad[6]= (q_2-p_1)^2-m_6^2.
\end {eqnarray}
Following the discussion given in~\cite{Ferroglia:2003yj}, we first combine
the propagators [1] and [2] with a Feynman-parameter $z_1$, the propagators
[3] and [4] with a Feynman-parameter $z_2$, and the propagators [5] and [6]
with a Feynman-parameter $z_3$. 
Then we combine the $q_1$ and $q_1-q_2$ propagators
with a parameter $x$.
After the $q_1$-integration we combine the residual propagators
with a parameter $y$ and carry out the $q_2$ integration. 
For the scalar integral $V^{\left(1\right)}_{222}$,
the resulting expression reads 
(we have changed $x\to1-x$ with respect to~\cite{Ferroglia:2003yj}):
\begin{eqnarray}
\label{v222}
&& V^{\left(1\right)}_{222} = 
\int^1_0\!\! dx\,dy\,dz_1\,dz_2\,dz_3\,
x\,y\,(1-x)\,(1-y)\,\,U^{-2}, 
\\[0.5cm]
&&U =  x\,y\,(1\!-\!x)(1\!-\!y)\,\xi
\,+\,y\,(1\!-\!x)\,\chi(z_1;p_2^2;m_1,m_2)
\nonumber \\
&& \hspace*{1cm}
\,+\,\,x\,y\,\chi(z_2;P^2;m_4,m_3)
\,+\,x\,(1\!-\!x)(1\!-\!y)\,\chi(z_3;p_1^2;m_5,m_6), 
\nonumber \\[0.3cm]
&& \xi = 
- \left[P^2\,(z_2\!-\!z_3)\,(1\!-\!z_1\!-\!z_2)
+ p_1^2\,(z_2\!-\!z_3)\,(1\!-\!z_1\!-\!z_3)
+ p_2^2\,(1\!-\!z_1\!-\!z_3)\,(1\!-\!z_1\!-\!z_2)\right],
\nonumber \\[0.3cm]
&& \chi(z;p^2;m,M) =
-p^2\,z\,(1\!-\!z) + m^2\,(1\!-\!z) + M^2\,z \, .
\nonumber
\end{eqnarray}
Here and in the following we use the short-hand notation
\begin{equation}
\int_0^1 \!\! dx\,dy \cdots dz_3 = 
\int_0^1 \!\! dx\,
\int_0^1 \!\! dy \cdots  
\int_0^1 \!\! dz_3 \, .
\nonumber
\end{equation}

\subsection{Non-planar diagram: two internal fermions}\label{V222a}

In Fig.~\ref{boscoll}(a) the following mass configuration of the 
$V_{222}$-family appears,
\begin{eqnarray}
m_1=m_5= 0,\quad m_2= m_e,\quad m_3= m_4= m_6= M_W,\qquad 
P^2=M_Z^2,\quad p_1^2= p_2^2= m_e^2 \, .
\end{eqnarray}
The divergence comes entirely from the fermion-photon interaction, 
so it is possible to set $p^2_1 = 0$.
Inserting these values into~(\ref{v222}) the resulting expression for the scalar integral $V^{\left(1\right)}_{222}$ reads
\begin{eqnarray}
V^{\left(1\right)}_{222} &=& 
\int^1_0\!\! dy\,dz_1\,dz_2\,dz_3\,y\,(1-y)\,I,
\quad {\rm with} \quad
I= \int^1_0\!\! dx\,\frac{x (1-x)}{\left[x\,(a x+b)+m^2_e\,c(x)\right]^2} \, ,
\label{V222div1}
\end{eqnarray}
where $a$, $b$ and $c\left(x\right)$ depend on $M_W$, $P^2$, 
and on the Feynman-parameters $y$, $z_1$, $z_2$, and $z_3$. 
In particular, $b$ is linear in $z_1$ and $z_3$,
\begin{eqnarray}
\nonumber b &=& \alpha z_1 z_3+\beta z_1+\gamma z_3+\delta, 
\quad \quad {\rm with}
\label{V222div1be}\\
\nonumber \alpha &=& M_Z^2\,y (1-y),\\
\nonumber \beta &=& -M_Z^2\,y (1-y) z_2,\\
\nonumber \gamma &=& (1-y)[ M^2_W-M_Z^2\,y\,(1-z_2) ],\\
\delta &=& y\,[ M^2_W - M_Z^2\,y\,z_2\,(1-z_2) ].
\end{eqnarray}
Moreover, the following relations hold,
\begin{eqnarray}
c(0) = y\,z_1^2,\qquad\quad
a+b = y\,[M_W^2-M_Z^2\,z_2(1-z_2)].\label{V222div1rel}
\end{eqnarray}
From (\ref{V222div1}) one can see that, for $m_e=0$, $I$ is divergent at $x=0$.
The divergence can be extracted as $\ln m^2_e $ in the following way:
\begin{eqnarray}
I= I_a + I_b=
\int_0^1\!\!\!dx\,\left\{ \bigg[
\frac{ x (1-x)}{[x\,(a x+b)+m^2_e\,c(x)]^2}-\frac{x}{[xb+m^2_e\,c(0)]^2}
\bigg]
+ \frac{x}{[xb+m^2_e\,c(0)]^2}
\right\}.
\end{eqnarray}
$I_a$ is not divergent and we are free to set $m_e=0$,
\begin{eqnarray}
I_a= 
\int^1_0\!\!dx\,\frac{1}{x}
\left[\frac{1-x}{\left(a x+b\right)^2}-\frac{1}{b^2}\right]
= -\frac{1}{b^2} \Big( \ln\frac{a+b}{b} + 1 \Big).
\end{eqnarray}
In $I_b$, the divergence can be extracted by integrating in $x$,
\begin{eqnarray}
I_b= - \,\frac{1}{b^2} \Big[ \ln m_e^2 + \ln c(0) - \ln b + 1 \Big].
\end{eqnarray}
Thanks to the linearity of $b$ with respect to $z_1$ and $z_3$, 
we can now perform the $z_1,z_3$ integrations  explicitly.
Taking into account (\ref{V222div1be}) and (\ref{V222div1rel}), the following types of integrals remain:
\begin{eqnarray}\label{skalInts}
\nonumber  
\int^1_0\!\!dz_1 \,dz_3\,\frac{1}{b^2} &=& 
\int^1_0 \frac{dz_1 \,dz_3}{(\alpha z_1 z_3+\beta z_1+\gamma z_3+\delta)^2}
\,=\, 
\frac{1}{d}\ln\left(\!1+\frac{d}{[\beta\delta]\,[\gamma\delta]} \!\right), 
\\[0.5cm]
%
\nonumber 
\int^1_0\!\!dz_1 \,dz_3\,\frac{\ln z_1}{b^2} 
&=& \frac{1}{d}\left\{
  \Li\left(\!-\frac{[\alpha\beta]}{[\gamma\delta]}\right)
- \Li\left(\!-\frac{\beta}{\delta}\right)
\right\}, \\[0.5cm]
%
\nonumber
\int^1_0\!\!dz_1 \,dz_3\,\frac{\ln b-1}{b^2} &=&
\frac{1}{d}\left\{
  \Li\left(\!\frac{d}{\alpha\,[\alpha\beta\gamma\delta]}\!\right)
- \Li\left(\!\frac{d}{\alpha\,[\beta\delta]}\!\right)
- \Li\left(\!\frac{d}{\alpha\,[\gamma\delta]}\!\right)
+ \Li\left(\!\frac{d}{\alpha\,\delta}\!\right)
\right.
\\
\nonumber &&
-\,\ln[\alpha\beta\gamma\delta]\,
   \ln\left(\!1\!-\!\frac{d}{\alpha\,[\alpha\beta\gamma\delta]}\!\right)
+  \ln[\beta\delta]\,\ln\left(\!1\!-\!\frac{d}{\alpha\,[\beta\delta]}\!\right)
\\
&&
\left.
+\,\ln[\gamma\delta]\,\ln\left(\!1\!-\!\frac{d}{\alpha\,[\gamma\delta]}\!\right)
-  \ln\delta\,\ln\left(\!1\!-\!\frac{d}{\alpha\delta}\!\right)
\right\},
\end{eqnarray}
where we have introduced $d = \alpha \delta-\beta \gamma$ and 
$[\alpha_1...\alpha_n]= \alpha_1+...+\alpha_n$.
It is important to note that the only denominator appearing in the result
($d$) is now multiplied by "regulator functions" which vanish when $d$ goes  
to 0.
The result is therefore smooth enough to be directly inserted into
(\ref{V222div1}), and the remaining $z_2,y$ integrations are done numerically.

The computation of the tensor integrals  $V^{(\mu,\mu\nu,\mu\nu\rho)}_{222}$ 
is processed in the same way. 
The new feature is the presence of $z_3$ to some power $m>0$ 
in the integrand. 
The previous formulae are now replaced by the following ones, 
where an integration by parts in $z_3$ is performed:

\begin{eqnarray}
\nonumber  
\int^1_0\!\!dz_1\,dz_3\,\frac{z_3^m}{b^2} &=&
\int^1_0\, dz_3\frac{m\,z_3^{m-1}}{d}
\ln\left(\!1+\frac{(1\!-\!z_3)\,d}
                {[\gamma\delta]\,([\alpha\gamma]z_3+[\beta\delta])}\!\right),
\end{eqnarray}

\begin{eqnarray}
\nonumber 
\int^1_0\!\!dz_1\,dz_3\,z_3^m\,\frac{\ln z_1}{b^2} &=&
\int^1_0\, dz_3\,\frac{m\,z_3^{m-1}}{d}
\left\{
  \Li\left(\!-\frac{[\alpha\beta]}{[\gamma\delta]}\right)
- \Li\left(\!-\frac{\alpha z_3+\beta}{\gamma z_3+\delta}\right)
\right\}, 
\end{eqnarray}

\begin{eqnarray}
\nonumber 
\int^1_0\!\!dz_1\,dz_3\,z_3^m\,\frac{(\ln b-1)}{b^2} &=&
\int^1_0\!\!dz_3\frac{m\,z_3^{m-1}}{d}
\left\{
  \Li\left(\!\frac{d}{\alpha\,[\alpha\beta\gamma\delta]}\!\right)
- \Li\left(\!\frac{d}{\alpha\,([\alpha\gamma] z_3+[\beta\delta])}\!\right)
\right.
\\
&&
\nonumber 
-\,\Li\left(\!\frac{d}{\alpha[\gamma\delta]}\!\right)
+  \Li\left(\!\frac{d}{\alpha(\gamma z_3\!+\!\delta)}\!\right)
-  \ln[\alpha\beta\gamma\delta]
   \ln\left(\!1\!-\!\frac{d}{\alpha\,[\alpha\beta\gamma\delta]}\!\right)
\\
&&
\nonumber 
+\,\ln([\alpha\gamma]z_3\!+\![\beta\delta])
   \ln\left(\!1\!-\!\frac{d}{\alpha([\alpha\gamma]z_3\!+\![\beta\delta])}\!\right)
\\
&&
\left. 
+\,\ln[\gamma\delta]
   \ln\left(\!1\!-\!\frac{d}{\alpha[\gamma\delta]}\!\right)
-  \ln(\gamma z_3\!+\!\delta)
   \ln\left(\!1\!-\!\frac{d}{\alpha(\gamma z_3\!+\!\delta)}\!\right)
\right\}.
\end{eqnarray}

\subsection{Non-planar diagram: four internal fermions}\label{V222b}

Fig.~\ref{boscoll}(b) corresponds to the configuration
\begin{eqnarray}
m_1= 0,\quad m_2= m_3= m_4= m_6= m_e,\quad m_5= M_Z\qquad 
P^2=M_Z^2,\quad p_1^2= p_2^2= m_e^2.
\end{eqnarray}
This is an example for a diagram with a double collinear divergence. In order
to regularize these divergences the masses of all internal electrons are kept
and $p_1^2=m^2_e=p_2^2$ is used. In the following the scalar integral
$V^{\left(1\right)}_{222}$ is considered; the tensor integrals can be
calculated as in section~\ref{V222a} .

We apply the same parametrization as in section~\ref{V222a} and change parameters according to $z_{2,3}\rightarrow 1-z_{2,3}$, $z_1 \leftrightarrow z_3$. The following expression is obtained, 
\begin{eqnarray}
V^{(1)}_{222} &=&
\int^1_0\!\! dy\,dz_3\,y\,(1-y)\,J,
\quad {\rm with} \quad
J = \int^1_0\!\!dx\,dz_1 \,dz_2\,
\frac{x\,(1-x)}{(\rho + m_e^2\,\sigma - i\,\delta)^2},
\label{v222fer4}
\\
\nonumber
\\
\nonumber
\rho &=&
-\, x\,\big\{\,
  y\,z_2\,(1-z_2) 
- (1\!-\!x)(1\!-\!y)[z_1 + y\,(z_1\!-\!z_2)(z_2\!-\!z_3)] 
\,\big\}\,M_Z^2 \, ,
\\
\nonumber
\sigma &=&
- \,x\,y\,(1\!-\!x)(1\!-\!y)(z_1\!-\!z_3)^2
+ (1\!-\!x)\,y \,z_3^2
+ x\,y
+ x\,(1\!-\!x)(1\!-\!y)\,(1\!-\!z_1)^2 \, .
\end {eqnarray}
Since $\rho$ is of the type $\rho= x\,( z_1\,A + z_2\,B + z_1\,z_2\,C)$,
one can easily see from eq.(\ref{v222fer4}) that, for $m_e=0$, 
$V^{(1)}_{222}$ diverges at $x=0$ and at $z_1=0=z_2$.
$J$ can be calculated in the following way:
\begin{eqnarray}\label{V222bgans}
J \,=\, J_a + J_b + J_c + J_d \,=\,
      [ J - J_x - J_z + J_{xz} ]
\,+\, [ J_x - J_{xz} ]
\,+\, [ J_z - J_{xz} ]
\,+\, J_{xz} \, ,
\end{eqnarray}
\vspace{-0.6cm}
\begin{eqnarray}
\nonumber
\,\,
J_x =
\int^1_0\!\!dx\,dz_1\,dz_2
\left.
\frac{x\,(1-x)}{(\rho + m^2_e\,\sigma)^2}
\right|_{x^2= x\,m_e^2= 0} ,
\qquad\,
J_z =
\int^1_0\!\!dx\,dz_1 \,dz_2
\left.
\frac{x\,(1-x)}{(\rho + m^2_e\,\sigma)^2}
\right|_{\!\!\!
\begin{array}{l}
\scriptstyle{z_1^2= z_2^2= z_1z_2= 0} \\
\scriptstyle{z_1\,m_e^2= z_2\,m_e^2= 0} 
\end{array} 
} ,
\end{eqnarray}
\vspace{-0.6cm}
\begin{eqnarray}
\nonumber
J_{xz} =
\int^1_0\!\!dx\,dz_1 \,dz_2\,
\left.
\frac{x\,(1-x)}{(\rho + m^2_e\,\sigma)^2}
\right|_{\!\!\!
\begin{array}{l}
\scriptstyle{x^2= z_1^2= z_2^2= z_1z_2= 0} \\
\scriptstyle{x\,m_e^2= z_1\,m_e^2= z_2\,m_e^2= 0} 
\end{array} 
} .
\end {eqnarray}
$J_a$ is finite for $m_e =0$, while $J_b$ and $J_c$ involve a single
divergence (at $x=0$ and $z_1=0=z_2$, respectively), 
giving raise to a simple logarithm in $m_e^2$. 
Finally, $J_d$ contains the double divergence at $x=0$ and $z_1=0=z_2$,
and we extract also a $\ln^2 m_e^2$ term accordingly.
Now we define:
\begin{eqnarray}
&a+b&= - y\,z_2\,(1\!-\!z_2)M_Z^2 - i\delta,
\\
\nonumber
&b&= 
\big\{ 
(1\!-\!y)[z_1 + y\,(z_1\!-\!z_2)(z_2\!-\!z_3)] - y\,z_2\,(1\!-\!z_2)]
\big\}M_Z^2 - i\delta ,
\\
\nonumber
&a_0+b_0&= - y\,z_2\,M_Z^2 - i\delta,
\\
\nonumber
&b_0&= 
\big\{ 
(1\!-\!y)[z_1 - y\,z_3\,(z_1\!-\!z_2)] - y\,z_2 
\big\}M_Z^2 - i\delta ,
\end{eqnarray}
\begin{eqnarray}
\nonumber
\begin{array}{ll}
A= (1\!-\!x)(1\!-\!y)(1\!-\!y\,z_3)M_Z^2 - i\delta ,
\qquad\quad&
B= -y\,[1 - (1\!-\!x)(1\!-\!y)\,z_3]M_Z^2 - i\delta ,
\\
A_0 \equiv A|_{x=0}= (1\!-\!y)(1\!-\!y\,z_3)M_Z^2 - i\delta ,
\qquad\quad&
B_0 \equiv B|_{x=0}= -y\,[1 - (1\!-\!y)\,z_3]M_Z^2 - i\delta ,
\end{array}
\end{eqnarray}
\begin{eqnarray}
\nonumber
\sigma_0
&=& 
\sigma|_{z_1=z_2=0}= 
\big[1\!-x\,(1\!-\!y)\big]\,\big[ x \!+\! (1\!-\!x)\,y\,z_3^2\big],
\\
\nonumber
\sigma_{00}
&=& 
\sigma|_{x=0}= \sigma|_{x=z_1=z_2=0}= 
y\,z_3^2 \, .
\end{eqnarray}
With these notations we obtain for $J_a$, setting $m_e=0$,
\begin{eqnarray}
\nonumber
J_a 
&=&
\int^1_0\!\!dx\,dz_1\,dz_2\,
\frac{1}{x}
\left[ 
  \frac{1-x}{(a\,x+b)^2}
- \frac{1}{b^2}
- \frac{1-x}{(a_0\,x+b_0)^2}
+ \frac{1}{b_0^2}
\right]
\\
&=& 
\int^1_0\!\!dz_1\,dz_2\,
\left[ 
  \frac{\ln b-\ln(a+b)-1}{b^2} 
- \frac{\ln b_0-\ln(a_0+b_0)-1}{b_0^2}
\right] .
\end{eqnarray}
The resulting expressions are of the form (\ref{skalInts}) and can be integrated in $z_1$ and $z_3$ accordingly. 
Since in this case both $\gamma$, $\delta$ and $d$ of (\ref{skalInts})
factorize a $z_2$, 
we get a factor $1$/$z_2$, that is not multiplied by regulator functions.
However, even if the integrals containing $1/b^2$ and $1/b_0^2$ are separately divergent, the combination of the two is finite, because they have the same behaviour around $z_2=0$.
Therefore, after applying (\ref{skalInts}), the contribution 
from $J_a$ to $V^{(1)}_{222}$ 
can be safely integrated numerically.

For $J_b$ we proceed in an analogous way integrating in $x$, and obtain
\begin{eqnarray}
\nonumber
J_b
&=&
\!\int^1_0\!\!dx\,dz_1\,dz_2
\left[
  \frac{x}{\left(x b + m^2_e\sigma_{00}\right)^2}
- \frac{x}{\left(x b_0 + m^2_e\sigma_{00}\right)^2}
\right]
\\
&=&
\!\int^1_0\!\!dz_1\,dz_2
\left[
  \frac{\ln b-\ln\sigma_{00}-1}{b^2} 
- \frac{\ln b_0-\ln\sigma_{00}-1}{b_0^2} 
- \ln(m^2_e)\left( \frac{1}{b^2}-\frac{1}{b_0^2} \right)
\right] .
\end{eqnarray}
Again, these terms are of the same form as in the expression for $J_a$ and can be integrated analytically in $z_1$ and $z_3$ and then numerically in $y$ and $z_2$.

To extract the logarithmic behaviour for $J_c$ we have to integrate in $z_1$
and $z_2$, yielding
\begin{eqnarray}\label{V222bzeh}
\nonumber
J_c
&=&
\int^1_0\!\!dx\,dz_1\,dz_2\,
\left\{
  \frac{x(1-x)}{[x(Az_1+Bz_2)+m_e^2\sigma_0]^2}
- \frac{x}{[x(A_0z_1+B_0z_2)+m_e^2\sigma_{00}]^2}
\right\}
\\
\nonumber
&=&
- \int^1_0\!\!dx\,\frac{1}{x} 
\left[
  \frac{(1-x)[\ln(A\!+\!B)-\ln A-\ln B - \ln x + \ln\sigma_0]}{AB}
\right.
\\
&&
\left. 
-\, \frac{\ln(A_0\!+\!B_0)-\ln A_0-\ln B_0 - \ln x + \ln\sigma_{00}}{A_0B_0}
+   \ln(m_e^2) \left( \frac{1-x}{AB} - \frac{1}{A_0B_0} \right)
\right] \!.\quad
\end{eqnarray}
The factors $y$, $(1-y)$, and $(1-x)$ in the denominators of this expression
are always cancelled by those present in the numerator. The overall factor
$1/x$ is again regularized
by the difference between the terms with $A$, $B$, $\sigma_0$ and those with $A_0$, $B_0$, $\sigma_{00}$. The remaining $y$ and $z_3$ integrations can be done numerically.

Finally, for $J_d$ we integrate in $x$, $z_1$ and $z_2$, yielding
\begin{eqnarray}
\nonumber
J_d
&=&
\int^1_0\!\!dx\,dz_1\,dz_2\,
\frac{x}{[x(A_0z_1+B_0z_2)+m_e^2\sigma_{00}]^2}
\\
&=&
\frac{1}{A_0 B_0}
\left[
  \Li\left( -\frac{A_0+B_0}{m^2_e\,\sigma_{00}} \right)
- \Li\left( -\frac{A_0}{m^2_e\,\sigma_{00}} \right)
- \Li\left( -\frac{B_0}{m^2_e\,\sigma_{00}} \right) 
\right] .
\end{eqnarray}
The dilogarithms can be expanded according to 
\begin{eqnarray}\label{L12ep}
\Li\left(-\frac{c}{m_e^2}\right) 
&=& 
- \frac{\pi^2}{6}
- \frac{1}{2}\,(\ln c-\ln m_e^2)^2 + \mathcal{O}(m_e^2),
\end{eqnarray}
giving rise to a term proportional to $\ln^2 m_e^2$. The resulting expression is suitable for numerical integration.

\section{Results}\label{Res}

\begin{table}[!htb]
\begin{tabularx}{16.cm}{l c  l}
\hline
parameter & \hspace*{10.cm} &value\\\hline
$M_W$ && $80.404$ GeV\\
$M_Z$ && $91.1876$ GeV\\
$\Gamma_Z$ && $2.4952$ GeV\\
$m_t$ &&  $172.5$ GeV \\
$\Delta\alpha\left(M^2_Z\right)$ && $0.05907$\\
$\alpha_s\left(M^2_Z\right)$ && $0.119$\\
$G_\mu$ &&  $1.16637 \times 10^{-5}$\\
$\overline{M}_W$ && $80.3766$ GeV\\
$\overline{M}_Z$ && $91.1535$ GeV\\
\hline
\end{tabularx}
\caption {\small Input parameters. $M_W$ and $M_Z$ are the experimental values
  for the $W$- and $Z$-boson mass, whereas $\overline{M}_W$ and
  $\overline{M}_Z$ are calculated quantities in the pole mass scheme. }
\label{tab:parameters}
\end{table}

\begin{table}[!htb]
\begin{tabularx}{16.cm}{c c c c c c}
\\
\\
\hline
$M_H\left[{\rm GeV}\right]$& &$\Delta\kappa^{(\alpha^2)}_{\rm ferm}\times
10^{-4}$ & $\Delta\kappa^{(\alpha^2)}_{\rm bos}\times 10^{-4}$ &
$\Delta\kappa^{(\alpha^2)}_{\rm ferm}\times 10^{-4}
$~\cite{Awramik:2006ar} & 
$\Delta\kappa^{(\alpha^2)}_{\rm bos}\times 10^{-4}$~\cite{Awramik:2006ar} \\
\hline
100  & &  1.07 &  -0.74 &  1.07 & -0.74  \\
200  & & -0.33 &  -0.47 & -0.32 & -0.47  \\
600  & & -2.89 &   0.18 & -2.89 &  0.17  \\
1000 & & -2.62 &   1.11 & -2.61 &  1.11  \\
\hline
\end{tabularx}
\caption {\small Bosonic two-loop corrections to $\Delta \kappa$ in
  comparison with the fermionic ones, and the results 
  from~\cite{Awramik:2006ar}.}
\label{tab:resultsbos}
\end{table}


\noindent 
In Tab.~\ref{tab:parameters} the set of input parameters
for our numerical evaluation
is listed. $M_W$ and $M_Z$ are the experimental values of 
the $W$- and $Z$-boson masses, which are the on-shell masses. 
They have to be converted to the values in the pole mass scheme~\cite{Freitas:2002ja},
labeled as $\overline{M}_W$ and $\overline{M}_Z$, 
which are used internally for the calculation.
These quantities are related via $M_{W,Z} = \overline{M}_{W,Z}+ \Gamma^2_{W,Z}/(2 \;M_{W,Z})$. 
For $\Gamma_Z$ the experimental value (Tab.~\ref{tab:parameters}) and for $\Gamma_W$ the
theoretical value has been used, {\it i.e.}  
$\Gamma_W = 
3 \;G_{\mu} M^3_W/\left(2 \sqrt{2}\pi\right) 
\left(1+ 2 \alpha_s\left(M_W^2\right)/\left(3 \pi\right)\right)$
with sufficient accuracy. 

Our results in terms of  $\Delta\kappa$ for the bosonic contributions are
shown in Tab.~\ref{tab:resultsbos}, in comparison with the fermionic ones. The
last two columns of Tab.~\ref{tab:resultsbos} contain the corresponding
results from~\cite{Awramik:2006ar}, which agree with ours.

One can see from Tab.~\ref{tab:resultsbos} that the bosonic corrections 
are basically of the same order of magnitude as the fermionic ones, 
as long as the 
$W$-mass is taken as a fixed input parameter.
The situation changes, however, 
when instead of the $W$-mass the 
Fermi constant is taken as an input parameter 
and $M_W$ is derived via (\ref{interdependence}). 
In order to identify the various sources of
the two-loop contributions to $\seff$, we expand
both quantities $M_W$ and $\kappa$ according to 
$\{M_W,\kappa\} = \{M_W,\kappa\}^{(\rm tree + \alpha)} +
\{M_W,\kappa\}^{(\alpha^2)} + {\cal O}(\alpha^3)$, yielding
\begin{eqnarray}
\seff 
&=&
  \seff^{({\rm tree} + \alpha)}
+ \Delta\seff(\Delta M_W^{(\alpha^2)})
+ \Delta\seff(\Delta \kappa^{(\alpha^2)})
+ {\cal O}(\alpha^3)
\\
\nonumber
&=&\!
\left( 1-\frac{(M_W^{({\rm tree} + \alpha)})^2}{M_Z^2} \right)\!
\kappa^{\!({\rm tree} + \alpha)}
- 2\,\frac{m_W}{m_Z^2}\,\Delta M_W^{(\alpha^2)}
+ \left(\! 1 \!-\! \frac{m_W^2}{m_Z^2} \!\right)\!\Delta\kappa^{(\alpha^2)}
+ {\cal O}(\alpha^3)
\end{eqnarray}

The first term represents the  one-loop result,
while the other two terms correspond to the electroweak two-loop
contributions to the shift in the $W$ mass, $\Delta M_W^{(\alpha^2)}$,
and to the shift in $\kappa$, $\Delta \kappa^{(\alpha^2)}$.
As can be seen in Tab.~\ref{tab:seff}, the fermionic corrections from $M_W$ 
dominate, while the contributions from $M_W$ and $\kappa$ in the bosonic
sector cancel to a large extent.

\begin{table}[!htb]
\centering
\begin{tabular}{c c c c c}
\hline
\noalign{\smallskip}
$M_H$ 
& 
$\Delta\seff(\Delta M_{W,\rm ferm}^{(\alpha^2)})$
&
$\Delta\seff(\Delta k_{\rm ferm}^{(\alpha^2)})$
&
$\Delta\seff(\Delta M_{W,\rm bos}^{(\alpha^2)})$
& 
$\Delta\seff(\Delta k_{\rm bos}^{(\alpha^2)})$
\\
\noalign{\smallskip}
\hline
\noalign{\smallskip}
100  &  93.89 &  2.38 &  1.93 & -1.65 \\
200  &  98.51 & -0.73 &  0.97 & -1.05 \\
600  & 106.89 & -6.43 &  0.19 &  0.40 \\
1000 & 105.36 & -5.83 & -1.16 &  2.47 \\
\noalign{\smallskip}
\hline
\end{tabular}
\caption{Fermionic and bosonic electroweak two-loop corrections to $\seff$.
$M_H$ is given in GeV; the values for $\Delta\seff$ 
have to be multiplied by $10^{-5}$.}
\label{tab:seff}
\end{table}

\begin{table}[!htb]
\begin{tabularx}{16.cm}{c c c c c c c}
\\
\\
\hline
$M_H\left[\mathrm{GeV}\right]$& &
$M_W^\mathrm{\rm ferm}\left[\mathrm{GeV}\right]$& 
$M_W^\mathrm{\rm ferm+bos}\left[\mathrm{GeV}\right]$& 
$\seff^\mathrm{ferm}$& $\seff^{\rm ferm+bos}$& 
$\seff^{\rm bos}\left[10^{-4}\right]$\\\hline
100  & & 80.3694&  80.3684&  0.231459 &  0.231461 &  0.02 \\
200  & & 80.3276&  80.3270&  0.231792 &  0.231792 &  0  \\
600  & & 80.2491&  80.2490&  0.232346 &  0.232352 &  0.06 \\
1000 & & 80.2134&  80.2141&  0.232587 &  0.232599 &   0.12\\
\hline
\end{tabularx}
\caption {\small $M_W$ and $\seff$  without (ferm) and with (ferm+bos) bosonic two-loop corretions. $\seff^{\rm bos}$ is the purely bosonic contribution to $\seff$.}
\label{tab:resultsbos2}
\end{table}

In Tab.~\ref{tab:resultsbos2}, the results for $M_W$, as taken from~\cite{Awramik:2006ar}, and
$\seff$ are shown with only the fermionic two-loop corrections 
($M_W^\mathrm{ferm}$ and $\seff^\mathrm{ferm}$) 
and with both fermionic and bosonic two-loop corrections 
($M_W^\mathrm{ferm+bos}$ and $\seff^\mathrm{ferm+bos}$) 
for various values of $M_H$. 
Besides the electroweak one- and two-loop corrections, the results for $\seff$ also contain the QCD-corrections of order $\mathcal{O}(\alpha\alpha_s)$ and $\mathcal{O}(\alpha\alpha_s^2)$~\cite{Chetyrkin:1995js}, as well as the leading three-loop corrections of order $\mathcal{O}(\alpha^2\alpha_s m_t^4)$ and $\mathcal{O}(\alpha^3 m_t^6)$~\cite{Faisst:2003px}.

The shifts in $\seff$ originating
from the full set of bosonic  two-loop contributions   
are displayed in the last column of Tab.~\ref{tab:resultsbos2}.
According to the cancellations mentioned above, 
they are much smaller than the corresponding shifts for 
a fixed value of the $W$ mass as an input parameter, which would result
from eq.~(\ref{kappa}) 
with the values of $\Delta\kappa$ given in Tab.~\ref{tab:resultsbos}.


\section{Conclusion}
\noindent In conclusion, we have completed the 
calculation of the electroweak bosonic two-loop
corrections to the effective leptonic mixing angle $\seff$. 
A comparison of our results with the ones in~\cite{Awramik:2006ar} 
was performed and full agreement was found. 
The bosonic corrections are 
basically of the same order of magnitude 
as the fermionic ones when the $W$-mass 
is taken as a fixed input parameter. 
If instead the Fermi constant is taken as input, with $M_W$ as a derived 
quantity, the bosonic corrections to $\seff$ are quite small.

In addition, the use of the general techniques of~\cite{Passarino:2006gv} to
treat IR-divergences provides a good test of their applicability to physical
processes. In particular we have shown that in this approach
the cancellation of the infrared poles can be verified analytically.
Moreover, our complete calculation of the two-loop electroweak corrections of
$\seff$ is the first physical application of the numerical methods of
\cite{Ferroglia:2003yj} showing that these methods are really applicable 
in practice for the calculation of  physical quantities. 
In particular within this approach massive two-loop vertices with many mass
scales are computed ``directly'', so not using mass expansions, which could
represent a good way to deal also with calculations beyond the Standard Model, 
where often many (unknown) massive particles come into the game.

\section*{Acknowldgement}
W.H. is grateful to the Institute of Theoretical Physics,
University of Vienna, where this paper was finalized
while he was Erwin Schr\"odinger visiting professor.


\end{document}